# LOTKIP: Low Overhead TKIP Optimization for Ad Hoc Wireless Network

**M. Razvi Doomun**\*
Computer Science and Engineering Department, Faculty of Engineering,
University of Mauritius, Reduit, Mauritius Tel: +230 781 8923; Fax: +230 465 7144
r.doomun@uom.ac.mu
\*Corresponding author

**K.M. Sunjiv Soyjaudah**
Electrical and Electronic Department, Faculty of Engineering,
University of Mauritius, Reduit, Mauritius Tel: +230 454 1041; Fax: +230 465 7144
ssoyjaudah@uom.ac.mu

**Abstract:**

Temporal Key Integrity Protocol (TKIP) is a provisional solution for Wired Equivalent Privacy (WEP) security loopholes present in already widely deployed legacy 802.11 wireless devices. In this work, we model and analyse the computational complexity of TKIP security mechanism and propose an optimised implementation, called LOTKIP, to decrease processing overhead for better energy efficient security performance. The LOTKIP improvements are based on minimising key mixing redundancy and a novel frame encapsulation with low overhead. We simulate and compare LOTKIP with baseline TKIP in terms of complexity and energy consumption for ad hoc wireless network security. From simulation results, we demonstrate that LOTKIP executes with lower computational complexity, hence, with faster encryption time and more energy-efficient.

**Biographical notes:**

**Mohammad Razvi Doomun** holds a B.Eng (Hons) in Electronic and Communication Engineering from University of Mauritius and an MSc in Multimedia Communications from University of Surrey, UK. He is lecturer in the Department of Computer Science and Engineering at University of Mauritius and currently doing his PhD in Wireless Security.

**K.M. Sunjiv Soyjaudah** received his BSc. (Hons.) degree in Physics from Queen Mary College, University of London in 1982, his M.Sc. degree in digital electronics from King's College, University of London in 1991 and his PhD degree from University of Mauritius in 1998. He is a chartered engineer and a member of the IEEE. He is presently Professor at the University of Mauritius. He has numerous journal and conference publications and his interests are communication theory, cryptography and wireless networks.

# 1.0 INTRODUCTION

As a security enhancement for IEEE 802.11 wireless networks, WiFi Protected Access (WPA) use Temporal Key Integrity Protocol (TKIP) mechanism for encryption. WPA is also optional in the new IEEE 802.11i security standard (referred as WPA2) [11]. TKIP mechanism is a provisional scheme used to strengthen security of IEEE 802.11 WLAN and is implemented through software advances. It reuses RC4 of Wired Equivalent Privacy (WEP) protocol as its core, but introduces upgrading in the areas of message integrity, IV creation, and key management and plays the part of a wrapper to increase the security of WEP [1]. However, TKIP security mechanism consumes precious CPU cycles in 802.11b wireless network devices as it incurs extra computation and communication overhead. For battery-powered and low processing capacity wireless devices, TKIP encryption/decryption operations cause extra delays in communication, decrease in effective bandwidth and increase energy consumption. The encapsulation process and message integrity check increase the size of transmitted packets, which in turn lower the effective bandwidth and increase the communication cost. Further, executing wireless security protocol, even including key exchanges, leads to extra network traffic. Hence, minimising security overhead and optimising power consumption are important challenges to wireless security design.

Supplementary power and resource utilization drain that TKIP security enhancements impose require research attention. There is a need for comprehensive quantitative security and complexity analysis of TKIP key mixing function and encapsulation process, even so cryptographic review thus far suggests it achieves its fundamental design goals. Overall processing limitations as well as energy consumption need to be alleviated with more efficient TKIP security implementations, which is especially desirable in battery-powered wireless devices for certain ad hoc wireless networks.

In this paper, we present a mathematical relationship model between power consumption and TKIP encryption complexity. A simple and efficient low overhead TKIP (LOTKIP) technique is also proposed as a trade-off between security overhead and power consumption which is suitable for ad hoc wireless network security. We consider TKIP key mixing function and RC4 encryption algorithms specified in [11] for our analysis and simulations. Energy consumption models of 802.11b wireless device operation and cryptographic algorithm processing are used to simulate and validate the performance and efficiency of LOTKIP for specific scenarios.

The rest of the paper is organized as follows: Section 2.0 gives an account of related work in the field of wireless security optimisation and energy consumption models. We present an overview of TKIP algorithm in Section 3.0. The TKIP complexity model is analyzed in Section 4.0. Then, we explain LOTKIP details in section 5.0. The simulations and performance of LOTKIP are discussed in Section 6.0, followed by concluding remarks in Section 7.0.

**2.0 RELATED WORK**

Most wireless security algorithms are designed based on models that do not take into account the security performance together with computational complexity and energy cost. In [16], a lightweight enhancement (LWE) to RC4 is proposed to operate in resource constrained wireless devices where 64-bit WEP is hardwired. The enhancement approach is based on derangement and complementation that use a block cipher mode of operation on top of 64-bit RC stream cipher to enhance security. LWE exponentially increase security strength with logarithmic expenditure of memory and power. State Based WEP presented by Srinivasan *et al.* [21] provides a strong, lightweight encryption scheme for battery-constraint wireless devices. Sate Based Key Hop saves significant processing power especially for packet sizes smaller than 200 bytes as would be seen in wireless networks by avoiding RC4 state initialization on every packet. It eliminates all the security issues with WEP using the existing hardware at a speed greater than WEP and Wi-Fi Protected Access (WPA). It has been noted that the energy consumption of most cryptographic algorithms increases in their software instantiations [5] [20]. Hence, bulk data encryption and message authentication algorithms for wireless security are among the dominant power sink in mobile wireless devices. Jones *et al.* [12] and Lettieri & Srivastava [15] have shown that one of the main causes of unnecessary energy consumption is security overhead and communication protocol over a wireless channel. Adaptability and optimization of security protocols have thus emerged as a key issue for security in ad hoc resource-constrained networks. Two principles suggested for achieving an energy efficient security system are to avoid unnecessary computations and reduce the amount/size of encrypted data transmission [10]. However, existing security protocols for 802.11 wireless networks limit the efficient use of wireless station/node resources by significantly increasing amount of overhead required to secure data communication and decreasing throughput. Complex cryptographic processing also increases the delay between data transmissions. With unoptimized security protocols the data rates of wireless links decreases due to additional traffic or larger encrypted packet size incurred for authentication or verification services. Essentially, security mechanisms increase overall power and energy consumption in wireless devices, since computationally complex encryption and decryption procedures require multiple arithmetic operations and more processing cycles. In [8], the authors have measured the actual energy consumption of 802.11 wireless network interfaces operating in ad-hoc network environments and showed that the amount of energy consumption is directly proportional to the size of data to be sent or received. The energy consumption models for data communication (transmission and reception) is expressed as linear equation of the form, $E = (m \times size) + b$, where the coefficients $m$ and $b$ depend on the type of communication, i.e., broadcast, unicast, or packet discarded, and can be determined empirically. Similarly, in [14], the energy consumption per packet of cipher function is also modelled to be almost linear as a function of: a fixed cost $(B)$ which is the energy overhead resulting from the computation required for key expansion process and is independent of the packet size, and a variable cost $(A)$ that is dependent on the packet size. The overall model is simplified as $E = B + xA$ for energy consumption per packet of size $x$ bytes. These approved general models have been useful in performance evaluation of energy consumption of wireless security protocols.

Several techniques have been investigated in [14] [5], to reduce energy consumption by limiting the duration transmission/reception of messages or designing more energy efficient idling techniques. Another active area of research interest is the optimization of security protocol efficiencies [7] [19] [13]. Prior work from Ganesan *et al.* [9] assesses the feasibility of different encryption schemes for a range of embedded architectures using execution time overhead measurements. Potlapally *et al.* [18] investigated energy consumption of different ciphers on the Secure Sockets Layer. Consequently, our work consolidates all earlier work on 802.11 wireless securities and adds to fill this void of, by investigating design of energy efficient and low overhead TKIP encryption in an ad hoc wireless network scenario.

## 3.0 TKIP ALGORITHM

In this section we review the basics of TKIP algorithm.

### 3.1 TKIP Basics

TKIP provides more security than WEP with no extra hardware. Based on specific redesigned attributes, TKIP algorithm fulfils the challenges of a higher security standard in the following ways: (i) Michael, a well-studied cryptographic message integrity code (MIC), is used for defeating forgeries. MIC is computed over the whole message and it also protects the source and destination address from falsification. (ii) A frame sequence numbering is used to defeat replay attacks. Out-of-order frames are flagged and replay attacks are mitigated. Attackers cannot capture valid encrypted traffic and re-transmit it at a later time. (iii) A per-frame key mixing function is employed to defeat weak key attacks. TKIP derives a unique RC4 key for each frame through key mixing process to mitigate attacks against weak WEP keys. The increased length of the IV makes it possible to generate a larger number of different keys. (iv) Lastly, a rekeying mechanism is included to defeat key collisions. TKIP is also equipped with key management operations. Another important constraint in the design of security mechanism for 802.11b and 802.11g mobile devices is the low-speed embedded CPU's which cannot support computationally intensive security operations. We node here the idea behind TKIP is compatibility with existing hardware while minimizing the impact of enhancement operations on the device performance.

**In Figure 1,** the TKIP encryption/decryption process between two wireless stations, A and B, is shown. At station A, a MIC is generated for the data and appended to the message which is fragmented if greater than MPDU size. Next, the MAC Header is added and the whole packet is encrypted. Before de-encapsulating a received MAC Protocol Data Unit (MPDU) at station B, TKIP extracts the TKIP sequence counter (TSC), WEP IV and Key ID from the packet. However, TKIP discards received MPDU that violates the sequencing rules, and otherwise uses the mixing function to reconstruct a WEP seed. The TKIP WEP seed is represented as a concatenation of WEP IV and RC4 key and passes on these with the MPDU to WEP engine for de-encapsulation/decryption. If integrity check value (ICV) test is successful, the implementation reassembles the MPDU into a MAC Protocol Service Unit (MSDU). Upon successful MSDU reassembly, the receiver station B performs MIC verification step by recomputing the MIC over the MSDU source address, destination address, and MSDU data (but not the MIC field), and bit-wise comparing the resultant MIC against the received MIC. Successful MIC verification means the MSDU can be delivered to the upper level. If two MICs differ in any bit position (interpreted as MIC failure), the receiver will discard the packet and will engage in appropriate countermeasures.

As specified in ref. [11], TKIP MPDU is extended by 4 bytes to accommodate the new Extended IV (ExtIV) field and the MSDU format is expanded by 8 bytes, to accommodate the new MIC field. The simplified layout of the encrypted MPDU format is depicted in **Figure 1**. When the MSDU-with-MIC cannot be encoded within a single WEP-encapsulated MDPU, it is fragmented into appropriately sized MPDUs. The 4 bytes of ExtIV are added after the existing IV/Key ID Field, i.e. the IV and Key ID of 4 bytes is retained in the form as defined with baseline WEP. If the ExtIV bit is '0' only the WEP style non-extended IV is transferred. When the ExtIV bit is set and the Extended IV field is supplied for TKIP, this indicates presence of extended mode to the receiver. The transmitting/receiving station keeps track of the IV value of to detect key exhaustion. As noted in **Figure 1**, the extended IV field is not encrypted [11]. All the MPDUs generated from one MSDU are encrypted under the same temporal key by TKIP.

TKIP employs non-reusable IVs as TKIP sequence counters (TSC) to prevent replay attacks [17]. With the same temporal or session key (TK), TSC is a monotonically increasing counter from 0x000 to 0xFFFF which starts from first packet transmission. The receiver rejects every packet that has a TSC less than or equal to the previous packet. When an IV sequence counter roll-over (0xFFFF --> 0x0000) is detected, the extended IV will be incremented.

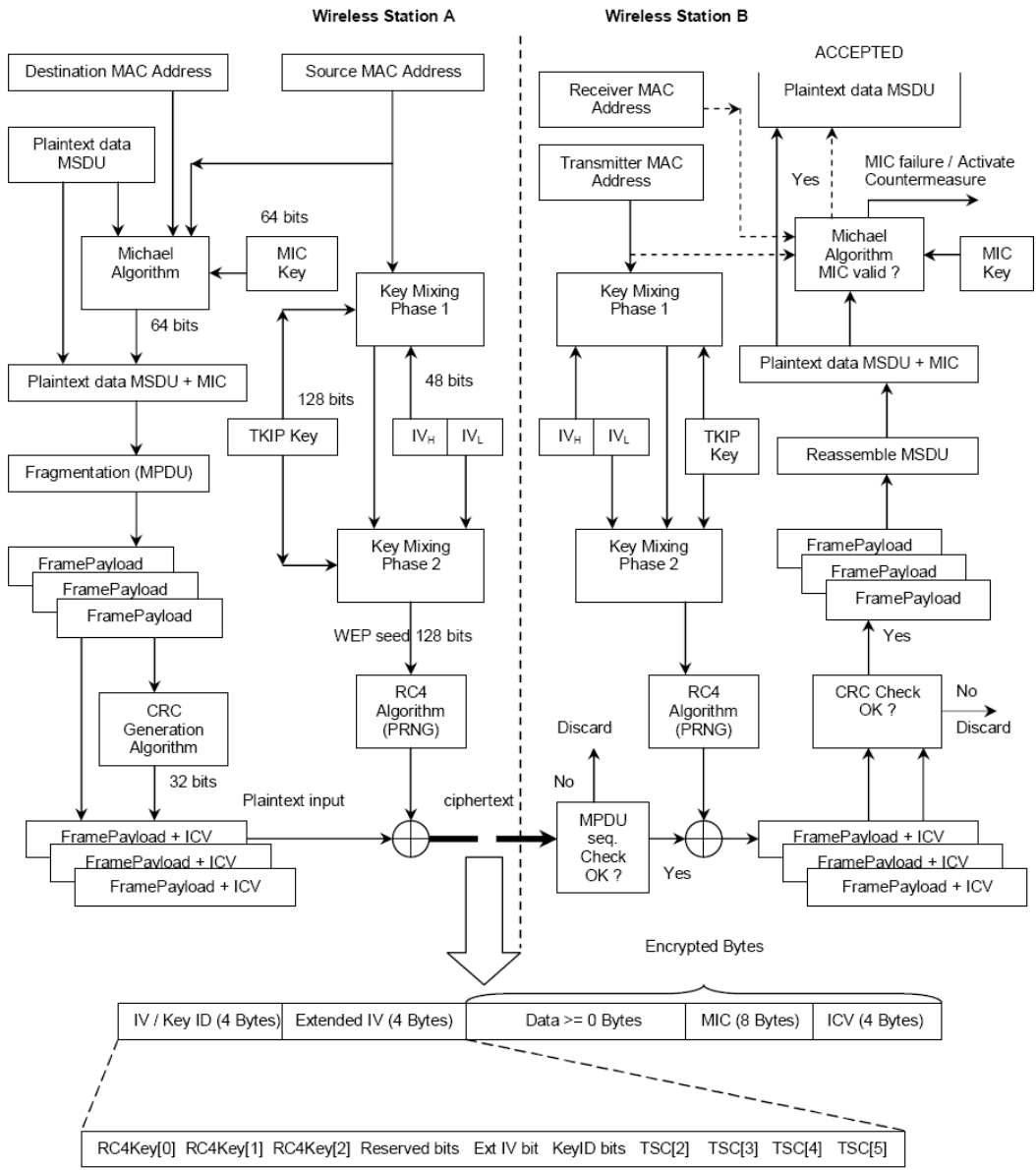

**Figure 1:** TKIP encryption/decryption and expanded MPDU structure

A total of 20 bytes headers are associated with TKIP IEEE 802.11 frame. The extended IV (4 bytes) and the MIC (8 bytes) amounts an overhead of 12 bytes. The setback of the TKIP sequence counter approach is related to IEEE 802.11 burst-acknowledgements, which indicates that up to 16 packets could be sent at once and then be acknowledged by one packet. The concept of replay window is used to monitor the counters. The receiver keeps track of the highest TSC and the last 16 TSC values and when a new frame arrives it checks and classifies it as one of the following types: ACCEPT if TSC is larger than the largest seen so far, or REJECT if TSC is less than the value of the largest 16, or adjust the WINDOW if TSC is less than the largest, but more than the lower limit. Consequently, the sequence counter memorizes the last 16 IV values to guarantee that all packets have been correctly received. Frames are fragmented in the band from 256 to 2346 bytes threshold with the purpose to increase the transfer reliability. Larger frame size increases likelihood of interference and therefore higher retransmission rate. Small frame rate gives excessive overhead during the transmission and throughput reduction.

### 3.1 TKIP Countermeasures

MIC detects active attacks (unlike WEP's Integrity Check Value (ICV)) and countermeasures are employed to prevent persistent message forgery attacks. However, TKIP still uses ICV in conjunction with the MIC to prevent false detection of MIC failures, and therefore thwart false countermeasure initiation. TKIP takes active countermeasures when two MIC failures are detected in less than one minute [11]. For MIC failure rate above one per minute, the station basically deletes all keys and attempts to reassociate (re-keying the connection) once more after a waiting period. However, this also implies that the attacker needs only two MIC-invalid packets per minute to completely prevent Wi-Fi users accessing the network. Although, normal network operation can resume at least 60s after the second MIC failure, to prevent this countermeasure from being used as a pedestal for a denial of service attack, the MIC is checked last in the TKIP de-encapsulation/decryption process. The risk of false alarms is minimized when MIC is verified after the CRC, IV and other checks have been performed [11]. Frames with invalid ICV and TSC are discarded before the MIC is verified.

Thus, ICV ensures that noise and transmission errors do not erroneously trigger TKIP countermeasures. Another countermeasure, when a new temporal key cannot be established before the full 16-bit space TSC is exhausted, then TKIP protected communications will cease. For key refresh failure, the implementation halts further data traffic until rekeying succeeds, or disassociates. Further, to strengthen the user authentication process, TKIP makes use of the 802.1x framework and the Extensible Authentication Protocol (EAP) as proposed in [11]. But even when addressing all known flaws of WEP, TKIP does not protect against denial-of-service (DoS) attacks, since the countermeasure can be used to launch a DoS attack on the network. DoS attack can exhaust resources such as bandwidth, memory, CPU, etc. Moreover, TKIP is designed in such a way that its security completely relies on the secrecy of all the packet keys [17].

TKIP stronger security comes at the cost of performance degradation, in terms of higher complexity and overhead. Key mixing operation is designed to put a minimum demand on the stations and access points, yet have enough cryptographic strength so that it cannot easily be broken. Key-mixing would be more CPU-intensive if not solved by a two-phase mixing process. Phase 1 key mixing is static and one-off with high 32 bits of the IV, and only changes every 64K packets. Phase 2 is executed on per-packet basic, but since the counter is predictable, phase 2 can be computed in advance while waiting for the next packet(s) to arrive at receiver. Computing a few mixed keys in advanced is a gainful approach to minimize decryption response time for strict time-constraint applications. Furthermore, in ad hoc IEEE 802.11 wireless networks, the link throughput can be improved with an optimized TKIP mechanism.

**4.0 COMPLEXITY ANALYSIS**

In this section we analyze and model the complexity of TKIP components.

**4.1 Michael Complexity Analysis**

Several viable implementations exist for secure message integrity checking, but the dilemma of the wireless LAN card is its limited finite computation capabilities. WPA uses cryptographic Message Integrity Code (MIC), called Michael, to support the ineffective Cyclic Redundacy Checksum (CRC) in WEP. Its characteristics conform to first generation Access Point devices having low 32-bit CISC power processors with limited MIP budget of 4 millions of instructions per second having to reuse existing WEP hardware [23]. MIC has three components: a secret authentication key (K), a tagging function, and a verification predicate. TKIP cannot use computationally intensive cryptographic methods over existing 802.11b WLAN hardware with low-power processor. Even if cryptographic computations are shifted to software level in clients, it is hard for resource-constrained devices to sustain heavy computations. Thus, Michael is chosen as usable method for computing MIC in TKIP. Compared to WEP, the overall TKIP is a costly process and could degrade performance at many access points by consuming every spare CPU cycle. MIC algorithm operates on MSDUs to reduce overhead as it is not necessary to append a MIC value to every MPDU fragment of the message. In distinction, TKIP encryption functions at the MPDU level. Michael key is 64 bits, represented as two 32-bit little-endian words *($K_0,K_1$)*. Given a message of the form *(Source_Adr||Destination_Adr||Data)*, it is partitioned into a sequence of 32-bit chunks and the last one is padded, if not full, with **0x5A** and enough zeros. Hence, the last word *$M_n$* is always zero and the second to the last word is always non-zero because of **0x5A** padding. An iteration is performed over the partitioned sequence of 32-bit words *$M_1, M_2, …, M_n$*, where n is the total number of words. In each step, *$M_i$* is mixed with the 64-bit key using XORs, rotations, bit swaps, and little-Endian additions, as shown in the code next. The tag is computed using the following simple iterative code structure:

```
Algorithm Michael (key = (K₀,K₁),        Algorithm b(L, R)
Message = (M₀, …, Mₙ)                    //rotates, little-Endian additions, bit swaps
    (L,R) ← (K₀,K₁)                          R ← R ⊕ (L <<< 17)
    for i = 0 to n – 1 do                    L ← (R+L) mod 2³²
        L ← L ⊕ Mᵢ                           R ← R ⊕ XSWAP(L)
        (L,R) ← b(L,R)                       L ← (R+L) mod 2³²
    return (L,R) as the message's tag        R ← R ⊕ (L <<< 3)
                                             L ← (R+L) mod 2³²
                                             R ← R ⊕ (L >>> 2)
                                             L ← (R+L) mod 2³²
                                         return (L,R)
```

The main operations in MIC algorithm are: Exclusive OR (XOR), two's complement addition (ADD), rotate left/right (ROT), discarding any bits of higher significance than *n* (MOD), i.e. mod ($2^{32}$) and given a 32-bit word, swap lower 16 bits and upper 16 bits (XSWAP). The computational effort required for MIC calculation of message (*M*) is a function of the number of 32-bit words message size (*M₀, …, Mₙ*) and the number of processing cycles required for performing all basic operations in MIC algorithm is expressed as:

$$T_{MIC\text{-}CALC} = (n \times (4\ byte\ XOR)) + [n \times \{(4 \times (4\ byte\ XOR)) + (4 \times (4byte\ ADD) + (4 \times (XSWAP)) + (22 \times bit\text{-}wise\ ROT) + (4 \times MOD)\}]$$

The following Boolean equivalence assumptions and simplifications are made which have negligible impact on the overall computation: ADD and MOD operations are approximated to XOR operation complexity, XSWAP and ROT are approximated to SHIFT, and 22 bitwise ROT is rounded to 3 bytewise SHIFT. Then, we have:

$$T_{MIC\text{-}CALC} = 52n(XOR) + 19n(SHIFT)$$

Considering $T_{and}$, $T_{or}$, and $T_{shift}$ to denote the numbers of processing cycles required for performing basic operations of a byte-wise AND, a byte-wise OR, and a byte-wise SHIFT respectively. The equation is:

$$T_{MIC-CALC} = 52n\ (2\ T_{and} + T_{or}) + 19n\ T_{shift},$$

where **n** is number of 32-bit words in message *M*.

Hence, the approximate number of processing cycles required for performing all operations in MIC is expressed as:

$$T_{MIC-CALC} = 104n\ T_{and} + 52n\ T_{or} + 19n\ T_{shift}$$

## 4.2 CRC Complexity Analysis

The fundamental mathematics behind CRC algorithm is polynomial division. The table look-up CRC algorithm method is used with pre-calculated CRC values as shown next:

**Table Method of CRC Algorithm:**
// Initialize the CRC register
1. XOR the CRC most significant byte with the incoming message byte
2. Use this byte to index into the 256 entry table
3. Shift the CRC register to the left by one byte
4. XOR the CRC register with the value indexed into the table
5. Continue with step 1 until no more message bytes are left
6. XOR the CRC register with the final XOR value

Applying a similar complexity analysis approach, the number of processing cycles, $T_{CRC-CALC}$, required for performing all basic operations in CRC for *m* bytes is expressed in general terms as:

$$T_{CRC-CALC} = (4m + 2)\ AND + (2m + 1)\ OR + m\ SHIFT + m\ MEM,$$

Considering $T_{and}$, $T_{or}$, $T_{mem}$ and $T_{shift}$ to denote the numbers of processing cycles required for performing basic operations of a byte-wise AND, a byte-wise OR, a byte memory access and a byte-wise SHIFT respectively. The approximate number of processing cycles required for CRC on *m* bytes is given as:

$$T_{CRC-CALC} = (4m + 2)\ T_{and} + (2m + 1)\ T_{or} + m\ T_{shift} + m\ T_{mem}$$

## 4.3 Key Mixing Complexity Analysis

For efficient computation, key mixing benefits from two phase operations as follows:

**Phase I Key Mixing**

Phase 1 mixes Temporal Key (128-bit *TK*) with upper IV part *$TSC_0$ …. $TSC_5$* (48-bit) and the transmitter's MAC address (48-bit *TA*), to create an intermediate key called phase 1 key (128-bit *TTAK*). For performance optimization, 80-bit intermediate key *TTAK* is normally cached for *($2^{16}$ – 1)* packets and recomputed only when the temporal key is changed or updated. By mixing the MAC address with *TK*, Phase 1 key mixing ensures that even if various stations use the same temporal key, still different key streams will be generated. The main operations used in phase 1 are: XOR denoted as $\oplus$ and S-box substitution (S) with all arranged as 16-bit values. The following standard algorithm of Phase 1 which appears in [11] is employed. Input parameters are 48-bit TA *$TA_0$ …. $TA_5$*, 128-bit TK *$TK_0$ …. $TK_{15}$*, and *$TSC_0$…. $TSC_5$*. Output is intermediate key *$TTAK_0$ …….. $TTAK_4$*.

```
PHASE1_KEY_MIXING (TA₀ …. TA₅, TK₀ …. TK₁₅, TSC₀…. TSC₅)
    PHASE1_STEP1:
        TTAK₀ ← MK16 (TSC₃, TSC₂)
        TTAK₁ ← MK16 (TSC₅, TSC₄)
        TTAK₂ ← MK16 (TA₁, TA₀)
        TTAK₃ ← MK16 (TA₃, TA₂)
        TTAK₄ ← MK16 (TA₅, TA₄)

    PHASE1_STEP2:
    For i =0 to PHASE_LOOP_COUNT – 1
    {
        j ← 2·(i & 1)
        TTAK₀ ← TTAK₀ + S[TTAK₄ ⊕ MK16( TK₁₊ⱼ, TK₀₊ⱼ)]
        TTAK₁ ← TTAK₁ + S[TTAK₀ ⊕ MK16( TK₅₊ⱼ, TK₄₊ⱼ)]
        TTAK₂ ← TTAK₂ + S[TTAK₁ ⊕ MK16( TK₉₊ⱼ, TK₈₊ⱼ)]
        TTAK₃ ← TTAK₃ + S[TTAK₂ ⊕ MK16( TK₁₃₊ⱼ, TK₁₂₊ⱼ)]
        TTAK₄ ← TTAK₄ + S[TTAK₃ ⊕ MK16( TK₁₊ⱼ, TK₀₊ⱼ)] + i
    }
End // S-box: an invertible non-linear substitution table
```

The function MK16 constructs a 16-bit value from two 8-bit inputs as MK16(X,Y) = (256*X) + Y. We deduce the computational operations for Phase 1 key mixing as:

$$T_{phase1\ keyMx} = 2570\ BytewiseXOR + \{LOOP\_NUM \times (2590\ BytewiseXOR + 10\ BytewiseSBT)\}$$

Assuming LOOP_NUM = L, then

$$T_{phase1\ keyMx} = 2570\ BytewiseXOR + 2590L\ BytewiseXOR + 10L\ BytewiseSBT.$$

Considering $T_{and}$, $T_{or}$, and $T_{mem}$ to denote the number of processing cycles required for performing basic operations of a byte-wise AND, a byte-wise OR, and a byte-wise substitution (SBT) from memory, respectively, we have:

$$T_{phase1\ keyMx} = 2570\ (2\ T_{and} + T_{or}) + 2590L\ (2\ T_{and} + T_{or}) + 10L(T_{mem})$$

And, if L = 8, we obtain:

$$T_{phase1\ keyMx} = 46580\ T_{and} + 23290\ T_{or} + 80\ T_{mem}$$

**Phase 2 Key Mixing**

Phase 2 key mixing takes as input, **TTAK** (80 bits) with **TK** (128 bits) and the last 16 bits of IV to generate a unique 128-bit RC4 key, also known as WEP seed. It employs S-box substitution, rotate operation and addition operation to generate the 128-bit per-packet RC4 key (PPK). The 128-bit per-packet RC4 key has an internal structure that must conform to the WEP specification for compatibility. In both phase 1 and phase 2 key mixing function, an S-box is used for non-linear substitution and the strength of the cryptosystem depends heavily on the quality of the S-box lookup table [17]. The WEP seed is represented as an array of 8-bit values, $WEPSeed_0 \ldots WEPSeed_{15}$. When the TSC space is exhausted, we can either replace **TK** with a new one or stop the communications. The 32 bits IV fed in phase 1 is changed whenever the 16 bits in phase 2 have been used up. Phase 2 makes use of a variable Per-Packet key [**PPK**] of 96 bits. It is represented as an array of 16-bit values: $PPK_0 \ldots PPK_5$ and the **PPK** values are mapped onto the **TTAK** values after iterations of a loop. The exclusive-OR operation [⊕], addition operation [+], AND operation [&], OR operation [ | ], and right-bit SHIFT operation [>>>] are used in the specification of Phase 2 [11] as given next.

**PHASE2 STEP1** creates a copy of TTAK and includes the TSC
$PPK_0 \leftarrow TTAK_0$
$PPK_1 \leftarrow TTAK_1$
$PPK_2 \leftarrow TTAK_2$
$PPK_3 \leftarrow TTAK_3$
$PPK_4 \leftarrow TTAK_4$
$PPK_5 \leftarrow TTAK_4 + MK16 (TSC_1, TSC_0)$

**PHASE2 STEP2** is a 96-bit mixing, using S-box and RotR1 function
For i = 0 to PHASE1_LOOP_COUNT
{
$PPK_0 \leftarrow PPK_0 + S[PPK_5 \oplus MK16( TK_1, TK_0)]$
$PPK_1 \leftarrow PPK_1 + S[PPK_0 \oplus MK16( TK_3, TK_2)]$
$PPK_2 \leftarrow PPK_2 + S[PPK_1 \oplus MK16( TK_5, TK_4)]$
$PPK_3 \leftarrow PPK_3 + S[PPK_2 \oplus MK16( TK_7, TK_6)]$
$PPK_4 \leftarrow PPK_4 + S[PPK_3 \oplus MK16( TK_9, TK_8)]$
$PPK_5 \leftarrow PPK_5 + S[PPK_4 \oplus MK16( TK_{11}, TK_{10})]$
$PPK_0 \leftarrow PPK_0 + RotR1 (PPK_5 \oplus MK16 (TK_{13}, TK_{12}))$
$PPK_1 \leftarrow PPK_1 + RotR1 (PPK_0 \oplus MK16 (TK_{15}, TK_{14}))$
$PPK_2 \leftarrow PPK_2 + RotR1 (PPK_1)$
$PPK_3 \leftarrow PPK_3 + RotR1 (PPK_2)$
$PPK_4 \leftarrow PPK_4 + RotR1 (PPK_3)$
$PPK_5 \leftarrow PPK_5 + RotR1 (PPK_4) + i$
}

**PHASE2 STEP3** applies last TK bits and assigns 24-bit IV
$WEPSeed_0 \leftarrow TSC_1$
$WEPSeed_1 \leftarrow (TSC_1 | 0x20) \& 0x7F$
$WEPSeed_2 \leftarrow TSC_0$
$WEPSeed_3 \leftarrow Lo8 ((PPK_5 \oplus MK16 (TK_1, TK_0)) >> 1)$
For i=0 to 5
{
$WEPSeed_{4 + (2.i)} \leftarrow Lo8 (PPK_i)$
$WEPSeed_{5 + (2.i)} \leftarrow Hi8 (PPK_i)$
}
Output $WEPSeed_0 \ldots WEPSeed_{15}$

Lo8 function refers to the least significant 8 bits of the 16-bit input value, Hi8 function refers the most significant 8 bits of the 16-bit value, RotR1 function rotates its 16-bit argument one bit to the right and MK16 constructs a 16-bit value from two 8-bit inputs as MK16 (X,Y) $\leftarrow$ (256* X) +Y.

The number of processing cycles for computational operations of phase 2 key mixing is:

$T_{phase2\ keyMx}$ = 9341 *BytewiseAND* + 4671 *BytewiseOR* + 12 *BytewiseSBT* + 12 *BytewiseROT*

For simplicity, a BytewiseMUL computation is computationally equivalent to 256 BytewiseADD, which is approximated to 256 BytewiseXOR. Again, assuming $T_{and}$, $T_{or}$, $T_{rot}$ and $T_{mem}$ denote the numbers of processing cycles required for performing basic operations of a byte-wise AND, a byte-wise OR, and a byte-wise SBT from memory respectively.

$$T_{phase2\ keyMx} = 9341\ T_{and} + 4671\ T_{or} + 12\ T_{mem} + 12\ T_{rot}$$

In baseline TKIP, Phase Loop Count value is chosen as 8, this maintains a balance between robustness in key mixing and complexity of key generation.

**Combining Both Key Mixing Computations**

**Case 1: Without Phase 1 Caching scheme**

Key mixing (phase 1 and phase 2) process generates key stream for 128 bits data encryption, and the sum of key mixing computation, without phase 1 intermediate key caching, is expressed as:

$$T_{both\ keyMx}\ /\ \text{per byte encryption} = 3495\ T_{and} + 1748\ T_{or} + 6\ T_{mem} + 12\ T_{rot}$$

**Case 2: With phase 1 Caching scheme**

The first 16 bytes encryption computation is:

$$T_{both\ keyMx}\ /\ \text{per byte encryption} = 3495\ T_{and} + 1748\ T_{or} + 6\ T_{mem} + 12\ T_{rot}$$

Subsequently, all next 16 bytes encryption computations incur:

$$T_{both\ keyMx}\ /\ \text{per byte encryption} = 584\ T_{and} + 292\ T_{or} + 1\ T_{mem} + 1\ T_{rot}$$

**4.4 RC4 Complexity Analysis**

Standard RC4 algorithm is analyzed to determine its computational workload. RC4 process consists of Key scheduling component (KSA) and Pseudo random generation module (PRGA) to generate the key stream to be XOR with plaintext stream.

### Key Scheduling Algorithm (KSA) of RC4

```
KSA_(K)
    For i = 0... N − 1
        S[i] = i
    i = 0
    j = 0
    Repeat N times
        i = i +1
        j = j + S[i] + K[i mod L]
        Swap(S[i]; S[j])
```

Where N= 256

Number of processing cycles for computation of KSA is:

$$T_{KSA} = 256 \times (3\ ADD + MOD + SWAP)$$

Approximating *ADD* and *MOD* to *XOR* computation complexity, we have:

$$T_{KSA} = 2064\ AND + 512\ OR + 256\ SWAP$$

Assuming $T_{and}$, $T_{or}$, and $T_{swap}$ denote the number of processing cycles required for performing basic operations of a byte-wise AND, a byte-wise OR, and a byte-wise SWAP from, respectively. Thus, we have:

$$T_{KSA} = 2064\ T_{and} + 512\ T_{or} + 256\ T_{swap}$$

### Pseudo Random stream Generation Algorithm (PRGA) of RC4

```
PRGA(K)
Initialization:
    i = 0
    j = 0
Generation loop:
    i = i + 1
    j = j + S[i]
    Swap(S[i]; S[j])
    Output P = S[S[i] + S[j]]
```

Number of processing cycles for computation of PRGA is:

$$T_{PRGA}\ /\ per\ byte = (3\ XOR + SWAP + SUB)$$

Using $T_{and}$, $T_{or}$, $T_{sub}$ and $T_{swap}$ to denote the number of processing cycles required for performing basic operations of a byte-wise AND, a byte-wise OR, a byte SUB and a byte-wise SWAP from, respectively, we obtain:

$$T_{PRGA} \text{ / per byte} = 6\ AND + 3\ OR + SWAP + SUB$$

$$T_{PRGA} \text{ / per byte} = 6\ T_{and} + 3\ T_{or} + T_{swap} + T_{sub}$$

**Overall RC4 complexity**

The final operation in RC4 is to output (M[k] *XOR* RC4Key) to obtained the cipher text, where *M[0, 1, ..., N-1]* is the input message consisting of *N* bits. This additional *XOR* operation is included to obtain the overall number of processing cycles to compute encryption a byte of data as follows:

$$T_{RC4} \text{ / per byte} = T_{KSA} + T_{PRGA} \text{ / per byte} + XOR \text{ /byte}$$

Total number of processing cycles for *m* bytes of data encryption is,

$$T_{RC4} = (2064\ T_{and} + 512\ T_{or} + 256\ T_{swap}) + \{m \times (8\ T_{and} + 4\ T_{or} + T_{swap} + T_{sub})\},$$

| Message size (M) in bytes | Number of Processing Cycles of TKIP Functions | | | | | | |
|---|---|---|---|---|---|---|---|
| | $T_{MIC-CALC}$ | $T_{CRC-CALC}$ | Case 1 $T_{both\ keyMx}$ | Case 2 $T_{both\ keyMx}$ | $T_{RC4}$ | Case 1 $T_{TKIP}$ | Case 2 $T_{TKIP}$ |
| 16 | 700 | 131 | 84176 | 84176 | 3056 | 88063 | 88063 |
| 32 | 1400 | 259 | 168352 | 85054 | 3280 | 173291 | 89993 |
| 48 | 2100 | 387 | 252528 | 85932 | 3504 | 258519 | 91923 |
| 64 | 2800 | 515 | 336704 | 86810 | 3728 | 343747 | 93853 |
| 80 | 3500 | 643 | 420880 | 87668 | 3952 | 428975 | 95763 |
| 96 | 4200 | 771 | 505056 | 88566 | 4176 | 514203 | 97713 |
| 112 | 4900 | 899 | 589232 | 89444 | 4400 | 599431 | 99643 |
| 128 | 5600 | 1027 | 673408 | 90322 | 4624 | 684659 | 101573 |

**Table 1:** Complexity decomposition of TKIP functions

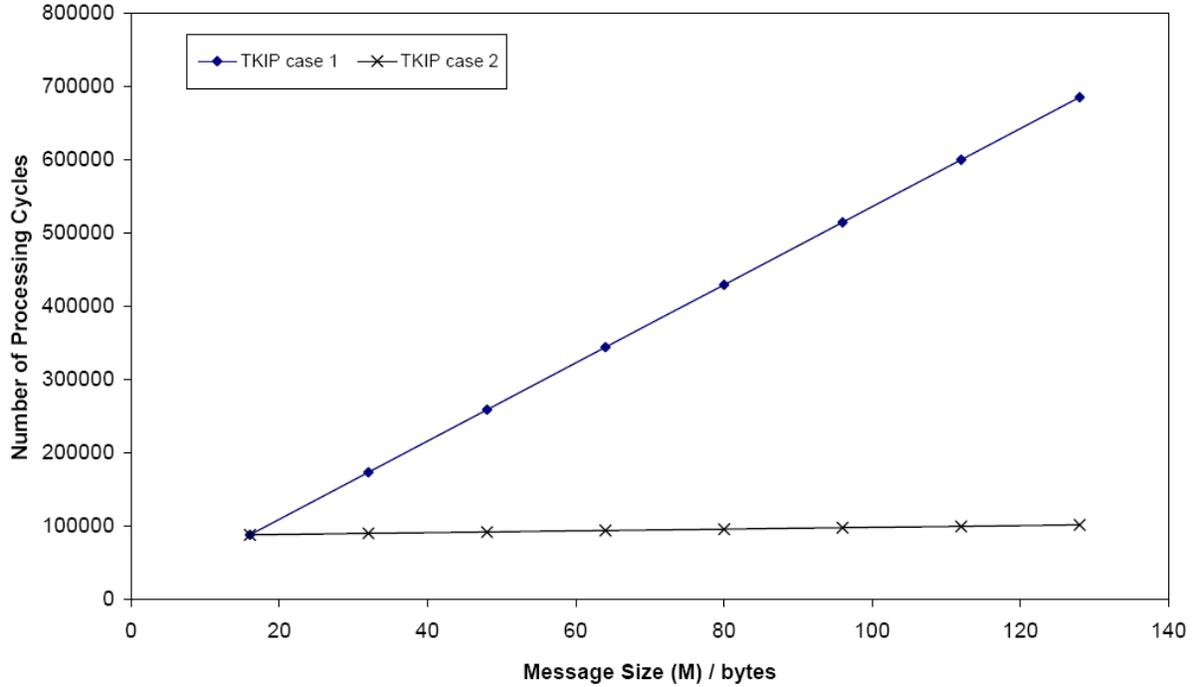

**Figure 2:** TKIP computational effort

As shown on **Table 1** and **Figure 2**, the complexity of TKIP is observed to increase linearly with larger messages size (*M*). Without phase 1 caching, overall TKIP complexity increases at a rate of 5386 processing cycles per byte encryption. When Phase 1 key mixing caching scheme is applied, the computational complexity of TKIP is much lower now, which also explains that key mixing is has the most operational workload of all the components of TKIP. Comparatively, MIC complexity percentage contribution for messages up till 128 bytes of data is (0.8 - 5.5%), CRC complexity part is (0.1 -1.0%), Key mixing complexity share is (88.9 - 95.6%) and RC4 complexity involvement is (3.5 - 4.6%). However, the larger the message size, the more complex it is to compute the MIC using Michael. For very large message sizes (of the order of Kilobytes) MIC function is expected to be the dominant complexity component.

**4.5 TKIP Packet Transmission Overhead**

| TKIP frame field | Purpose | Size (Bytes) |
|---|---|---|
| IV and Key ID | WEP seed, Key ID and partly used for key mixing | 4 |
| Extended IV | Extend the IV for sequence numbers and key mixing | 4 |
| MIC | Michael hash output | 8 |
| ICV | The WEP CRC checksum for integrity | 4 |
| *Overall TKIP packet encapsulation overhead* | | 20 |

**Table 2:** TKIP packet overhead field

TKIP does extra computation for per packet key generation. MIC check accounts for additional computation and increases the payload by 8 bytes. WEP appends only 8 bytes extra (i.e. 3 bytes for IV, 6 bits zero reserved, 2 bits key ID and 4 bytes ICV) in an IEEE 802.11 frame, whereas there are a total of 20 bytes (i.e. 4 bytes IV and key ID, 4 bytes extended IV, 8 bytes MIC and 4 bytes ICV) associated with TKIP in an IEEE 802.11 frame, as summarized in **Table 2**. Hence, TKIP extends the total length of a WEP encrypted MPDU by 12 bytes. IEEE 802.11 MSDU maximum size is explicitly 2304 bytes. The maximum MAC header and FCS overhead is 34 bytes, but only frames between access points over a wireless distribution system use all MAC header fields.

**4.6 TKIP Key Managment**

For key refresh mechanism, TKIP deals with three types of keys that are hierarchical [11] [23]: Master key (MK), Key Encryption key (KEK) and Temporal key (TK). Initially, a MK is exchanged among workstations through 802.1x authentication servers. MK is directly related to authentication and is used for secure distribution of key streams, i.e. it is created after a successful authentication and is related to one session only. Secondly, a pair of KEK is securely distributed between the authentication server and the wireless station via the AP using the MK. One KEK is needed to encrypt distributed keying material, i.e. temporal keys, while a second KEK serves to protect re-key messages from forgery. The station and the access point then generate a separate pair of TK for each direction of transmission in an association. Each pair of TK consists of a 128-bit data encryption key and a second 64-bit key for data integrity. These keys are identified by a 2-bit key ID. New TKs are always created with the first connection or re-establishment of connection.

In wireless ad hoc network, a key management system is also needed to distribute authentication and encryption keys to stations securely. The absence of an Access Point makes key management and distribution a challenging task. For the functionality of LOTKIP in ad hoc network, we assume that it is possible to set pre-shared key among the wireless stations, i.e. keys are distributed to the devices participating in the communication beforehand. This static approach will work if all the devices sharing the key are known when the ad hoc network is setup. Otherwise, cluster architecture is used for proactive key sharing when new stations authentication with the cluster head of the ad hoc wireless network. Any secure key distribution and authentication schemes in [2] [4] can be used.

**5.0 LOW OVERHEAD TKIP**

In baseline TKIP, phase 2 key mixing function reuses the intermediate 80-bit TKIP mixed Transmit address and key (TTAK) or (P1K) for $2^{16}$ consecutive MAC Protocol Data Units (MPDUs) during the same secure session. These MPDUs are encrypted with the same 32-bit upper Initialization vector (IV) part, Temporal Key (TK) and Transmitter Address (TA). Hence, caching of 80-bit TTAK it conventionally used. Since the knowledge of 32-bit high IV and the future sequence of 16-bit low IV is also known to the receiver after first packet tarnsmission, we modified the packet format to send the full 48-bit extended IV for the first packet initially and remove this redundancy in other successive packets. Therefore, in Low Overhead TKIP (LOTKIP) frame format, the redundant 4-bytes extended IV is removed from the packet load for packets ranging from the $2^{nd}$ to the $(2^{16}-1)^{th}$ packet. With this scheme, if the first packet is correctly received in an error free channel, decryption keys can be pre-computed in advance. **Figure 3** illustrates the packetization choice for LOTKIP mechanism. For wireless channel with frequent packet losses, a possible solution is to periodically include and send the 4-bytes extended IV (act as a refresh after every *K* MPDUs).

An acknowledgement (ACK) is sent back by the receiver to confirm LOTKIP packet delivery. Generally, the unicast packet of IEEE 802.11 does not support reordering, and the sender will only continue transmitting packets when it receives ACKs and packets with larger sequence number cannot overtake the packet with smaller sequence number. However, when wireless channel conditions degrade during transmission, the ACK message may not be received due to LOTKIP packets loss or ACK packet loss. An adaptive transmission control can be used, such that in the case of lost packets, data packet transmission stops and the source instead sends probes (short packets). Once acknowledgement of a probe is received, the sender resumes to normal transmission again with an initial LOTKIP packet *type A*, and then continues with LOTKIP packet *type B*. Further, LOTKIP encapsulation uses special flag bits for specific control purpose. The LOTKIP is particularly applicable for ad hoc wireless networks with short transmission range and low interference.

Next, in LOTKIP the MIC also includes the IV part in its tag computation. The transmitter computes a keyed cryptographic message integrity code over the MSDU source and destination addresses, the priority bits, the MSDU plaintext data and the 48-bit Extended IV also. LOTKIP appends the computed new MIC to the MSDU data prior to fragmentation into MPDUs. The receiver obviously has to verify the MIC after decryption with Extended IV, ICV checking, and reassembly of the MPDUs into an MSDU. Invalid MIC means discarding of corresponding MSDUs, and this defends against forgery attacks and replay attacks.

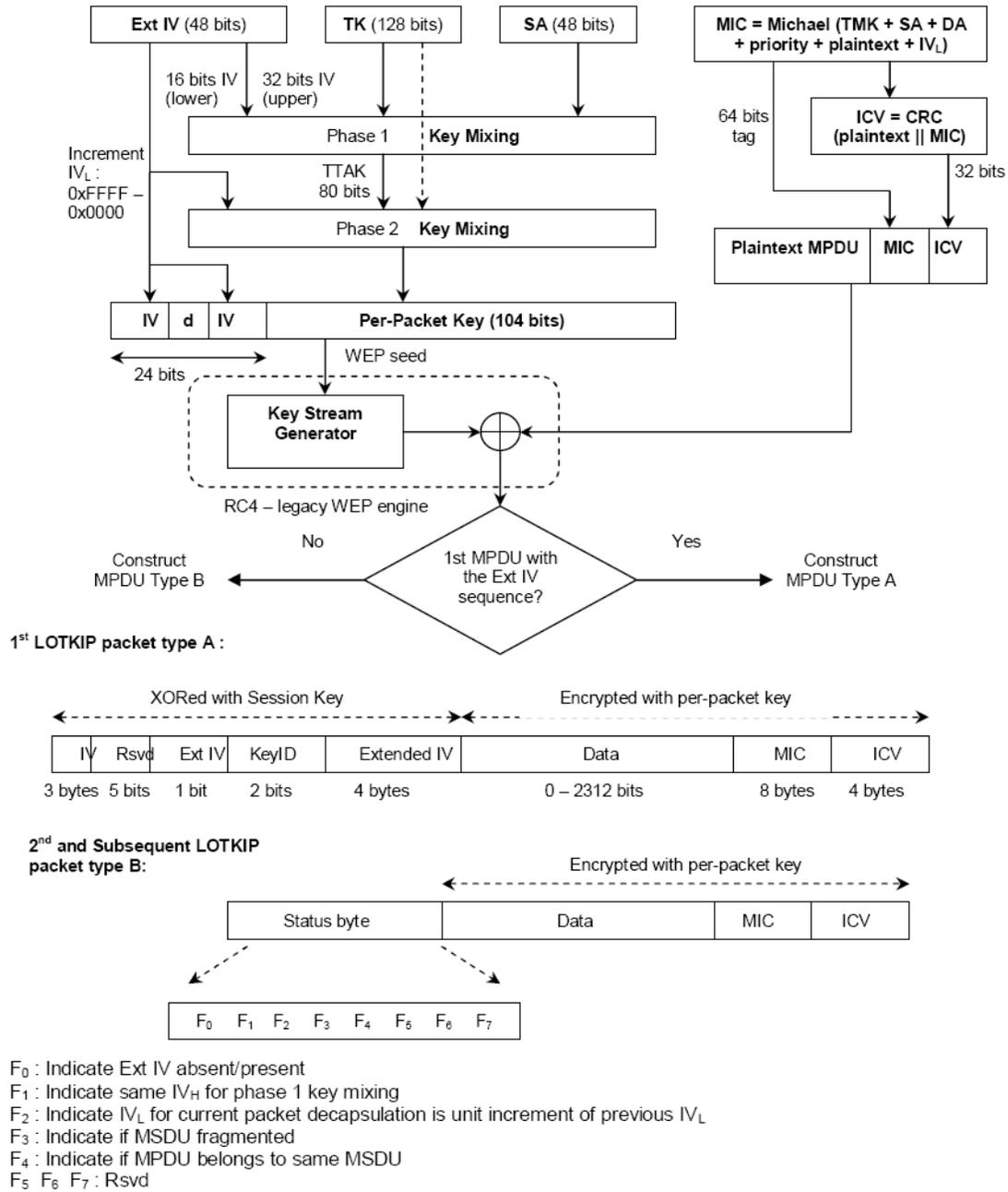

**Figure 3:** LOTKIP procedure and MPDU format

In the traditional baseline TKIP approach, the key reason why the IV is transmitted in the clear is because the 802.11 standard assumes that an adversary does not gain any useful information from its knowledge. The IV is meant to introduce randomness to the key, and appending the clear IV in the transmitted packet helps the receiver to decrypt the information sent from the transmitter station. However, it has been proved that various types of attacks are possible using the IV knowledge as described in [22] [23] [3]. The proposed LOTKIP heals this problem, as well minimizes packet overhead and thus potentially saving on transmission energy.

## 6.0 SIMULATION AND EVALUATION

### 6.1 Energy Consumption Models

802.11 WLAN device usually operate in one of the following modes: Transmit or Receive or Idle. When data bits are transmitted on the channel, the power consume is $P_T$ watts. When data bits are received and processed from the channel, the power consumed is $P_R$ watts. When the wireless card is idle, no transmission and reception of bits, the power spent is $P_I$ watts. Assuming the time length for which the WLAN card is in transmit, receive and idle mode are $T_T$, $T_R$ and $T_I$ respectively, the total energy consumed will be $(P_T \times T_T) + (P_R \times T_R) + (P_I \times T_I)$. The highest power is consumed in the transmit mode when transmitting packets or forwarding a packet in a multi-hop ad-hoc network. In the idle mode, a WLAN device is required to sense the medium and will be omitted in our simulation as it does not impact on our comparative analysis. We apply a linear energy consumption model for Lucent IEEE 802.11b wireless card operations as given in Feeney *et al*. [8]. According to this model, the energy consumption (E) in IEEE 802.11b networks is associated with the size of sent packets: *E = (a × Size) + b*, where *a* is the energy consumption per byte, and *b* is the overhead for sending a packet. The linear transmission and reception energy models for point-to-point transmission/receptions using IEEE 802.11b devices given by: *T_Energy = 431 µJ + 0.48 µJ/bytes;* and *R_Energy = 316 µJ + 0.12 µJ/bytes,* respectively [8].

## 6.2 TKIP Energy Model

Using the experimental results in [16], TKIP algorithm consumes *26804 µJ* per *256* byte frame on a 1.8 GHz intel© P-4 processor, 512 MB RAM. For the functionality of our energy simulation, we assume that *AND*, *OR*, *SHIFT*, *MEM*, *ROT* and *SWAP* basic operations can be each executed in one CPU cycle. Thus, using the complexity model in section 4, TKIP (without key caching) will encrypt *256* bytes using *1356683* processing cycles. One CPU processing cycle execution approximately costs *0.0198 µJ*. Simplifying the TKIP computational complexity equations, the TKIP energy model is given as follows:

**Case 1 (without key caching):**

$TKIP\_Energy = (175n + 5283m + 2835) \times 0.0198 \, \mu J$,

where $n = (m/32)$, $n \geq 1$, $m$ is the number of bytes encrypted.

**Case 2 (with key caching):**

1st Packet (*16 bytes*):
$TKIP\_Energy = (175n + 5283m + 2835) \times 0.0198 \, \mu J$,

where $n = (m/32)$, $n \geq 1$, $m$ is the number of bytes encrypted in first packet.

Subsequent Packets:
$TKIP\_Energy = (175n + 1764m + 2835) \times 0.0198 \, \mu J$,

where $n = (m/32)$, $n \geq 1$, $m$ is the number of bytes encrypted after first packet.

## 6.3 Network Model

We evaluate and compare the performance of the baseline TKIP and LOTKIP schemes in randomly-generated network topologies using 49 static wireless stations that are either transmitters or receivers are randomly placed within a (500 m x 500 m) flat area. The quasi unit-disk graph (*Quasi-UDG*) model is used to simulate the non-uniform characteristics of wireless networks [6]. A *Quasi-UDG* with transmission range parameters $R$ and *Quasi-UDG* factor $\alpha$ ( $0 \leq \alpha \leq 1$) over a set of positions in the network is defined as follows: For any two ad hoc wireless stations at positions $u$ and $v$ in the network with inter-station distance $|uv|$: if $|uv| \leq \alpha R$ then a link exists between wireless station at $u$ and $v$ in the network; if $|uv| > R$ then stations at $u$ and $v$ are not within direct transmission/reception range; and if $\alpha R < |uv| \leq R$ then stations at $u$ and $v$ is connected probabilistically. We simulate 100 different scenarios of randomly selected transmitter-receiver pairs among the 49 nodes in the ad hoc wireless network and the results are averaged. In this simulation, we assumed the receiver node cannot be compromised and there is an existing key management system as described previously.

**6.4 Results and Discussions**

The maximum IEEE 802.11 MSDU size is 2312 octets before the frame body is encrypted. Frames are fragmented and tested from 256 to 2312 bytes length. 10,000 packets of size ($P$) are transmitted from randomly selected sources and destinations in the 49 nodes ad hoc network. End-to-end baseline TKIP (without key caching) encryption/decryption is applied to secure communication. The simulation is run for grid distribution and random distribution ad hoc nodes. 20 bytes header is appended to the packets $P$ to form the TKIP encapsulated frames. The total energy consumption by all nodes in the network (Network Energy Consumption/Joules) is plotted in **Figure 4**. The dotted lines represent the results for random ad hoc wireless network.

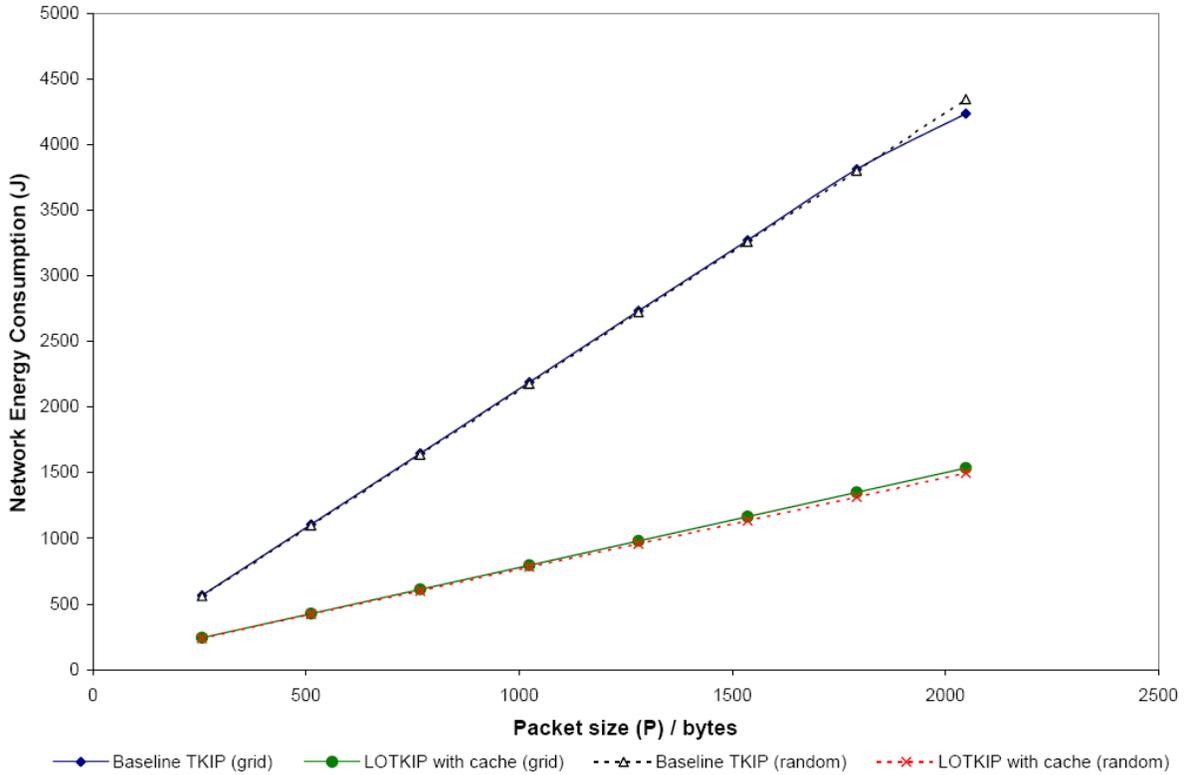

**Figure 4**: Network energy cost comparison

Dividing the Network Consumption Energy by the number of nodes/stations in the network, e.g. 49 nodes in our case, gives the average energy spent (Joules) per node/station. For brevity, we do not show the graph of average energy spent per node, since the overall trend is similar to **Figure 4**. As shown in the same figure, the network energy consumption increase linearly with larger packet size, assuming all other simulation parameters remain constant. The energy efficiency factor gained by LOTKIP is given as $\{2.33 + (P \times 0.00028)\}$, where P is the packet size ($256 \leq P \leq 2048$ bytes). The network energy consumption of baseline TKIP augments at a faster rate compare to LOTKIP as the packet size gets bigger. Therefore, LOTKIP could be a better scheme to transmit larger packets securely at lower network energy cost. The simple and efficient optimization scheme in LOTKIP shows lower overhead and energy saving. Since all packets depend of the first LOTKIP packet which contained the IV part, this packet is of high importance. In our simulation we assumed that this critical packet is not lost or intercepted by adversaries. In real network transmissions, the first LOTKIP packet can be treated in priority and delivered securely by VNP tunneling techniques.

However, we have not studied the energy consumption of WLAN attacks, such as denial of service attacks, related security countermeasures of LOTKIP. One approach to measure the efficiency of LOTKIP, and study how to minimize energy drain due to wireless LAN attacks, is to gauge its resilience to cryptanalysis attacks, such as brute-force attack, differential cryptanalysis and linear cryptanalysis.

## 7.0 CONCLUSION

TKIP encryption/decryption is one of the persistent overhead in wireless security for the entire communication session. In this paper, we have described, LOTKIP, a simple and optimized wireless security protocol that carries out power-efficient encryption and decryption. A mathematical complexity model of TKIP has been derived in terms of processing cycles of its basic operations and adapted to study the performance of LOTKIP. LOTKIP decreases complexity of wireless encryption while making LOTKIP frame transmission energy efficient.

**Acknowledgements**

The authors thank anonymous reviewers for their comments to improve the paper.